\documentclass[runningheads]{llncs}

\usepackage{graphicx}
\usepackage{enumerate}
\usepackage{hyperref}

\usepackage{comment}
\usepackage{enumitem}
\usepackage{amsmath}
\usepackage{amssymb}
\usepackage{algorithm}
\usepackage{algpseudocode}
\usepackage{subcaption}

\usepackage{array}

\usepackage{pdflscape}

\usepackage{longtable}

\usepackage[table,xcdraw]{xcolor} 
\usepackage{colortbl} 

\usepackage{setspace}

\begin{document}

\title{Federated Learning: An approach with Hybrid Homomorphic Encryption}
\titlerunning{Federated Learning: An approach with Hybrid Homomorphic Encryption}

\author{Pedro Correia\inst{1}\orcidID{0009-0009-4816-1755} \and Ivan Silva\inst{2}\thanks{Corresponding author: ivcsi@isep.ipp.pt}\orcidID{0009-0009-8480-9352} \and Ivone Amorim\inst{2}\orcidID{0000-0001-6102-6165} \and Eva Maia\inst{2}\orcidID{0000-0002-8075-531X} \and Isabel Praça\inst{2}\orcidID{0000-0002-2519-9859}}

\authorrunning{P. Correia et al.}

\institute{Faculty of Engineering, University of Porto, Porto, Portugal \\ 
\and GECAD, ISEP, Polytechnic of Porto, 4249-015, Porto, Portugal
}

\maketitle             

\begin{abstract}
Federated Learning (FL) is a distributed machine learning approach that promises privacy by keeping the data on the device. However,
gradient reconstruction and membership‑inference attacks show that model
updates still leak information.  Fully Homomorphic Encryption (FHE) can address those privacy concerns but it suffers from ciphertext expansion and requires prohibitive overhead on resource-constrained devices.  We propose the first {Hybrid Homomorphic Encryption}~(HHE)
framework for FL that pairs the PASTA symmetric cipher with the BFV FHE scheme.
Clients encrypt local model updates with PASTA and send both the lightweight ciphertexts and the PASTA key (itself BFV‑encrypted) to the server, which performs a homomorphic evaluation of the decryption circuit of PASTA and aggregates the resulting BFV ciphertexts.
A prototype implementation, developed on top of the Flower FL framework, shows that on independently and identically distributed MNIST dataset with 12 clients and 10 training rounds, the proposed HHE system achieves 97.6\% accuracy, just 1.3\%  below plaintext, while reducing client upload bandwidth by over 2,000$\times$ and cutting client runtime by 30\% compared to a system based solely on the BFV FHE scheme. However, server computational cost increases by roughly 15621$\times$ for each client participating in the training phase, a challenge to be addressed in future work.

\end{abstract}

\section{Introduction}
Federated learning~(FL) is a distributed machine learning approach that allows a group of clients, such as Internet of Things (IoT) devices, hospitals, or healthcare providers, to collaboratively train a global model without sharing their sensitive data. In typical FL setups, each client trains a local model using its own data and shares only the model parameters with a central server. The server then coordinates the training process by aggregating the received updates to form an improved global model. Once updated, the global parameters are sent back to the clients, allowing them to continue local training~\cite{mcmahan2023communicationefficientlearning}. Since raw data never leaves the device, FL is often described as ``privacy-preserving''~\cite{aledhariFederatedLearningSurvey2020}. However, recent work has shown that local model updates can still leak private information, enabling attacks such as gradient-based data reconstruction, membership inference, and property inference~\cite{10089719,10.1007/978-981-97-5603-2_36,NIU2024404,Yu2023}.
To address these issues, several techniques have been integrated into FL systems, such as Differential Privacy~(DP), Homomorphic Encryption~(HE), and Multi-Party Computation~(MPC)~\cite{stanSecureFederatedLearning2022}. However, these techniques also present practical limitations. DP adds noise to the training process.
MPC protocols often require multiple rounds of interaction and tend to scale poorly beyond a few hundred clients. 
HE allows servers to compute directly on encrypted data, offering strong and quantifiable privacy guarantees. Unfortunately, ciphertext expansion and the high algebraic depth of HE operations make pure HE-based FL much less efficient
than its plaintext counterpart~\cite{zhangHomomorphicEncryptionBasedPrivacyPreserving2023,jin2023fedmlhe}. These limitations make such approaches prohibitive for deployment in resource-constrained environments, motivating the Hybrid Homomorphic Encryption~(HHE) architecture introduced in this work.
HHE is a promising line of work that combines a lightweight \emph{symmetric} cipher with a HE one~\cite{abdinasibfarHHELandExploring2025}. In this type of scheme, data is first encrypted using a symmetric cipher, and sent to the server together with the symmetric key encrypted using HE. The server then does a homomorphic evaluation of the symmetric decryption circuit using the encrypted key and ciphertext, converting the symmetric ciphertext into a homomorphic one. This allows the server to perform computations on the encrypted data, ensuring that raw data remains protected~\cite{dobraunigPastaCaseHybrid2021}. Although this approach introduces additional computational cost on the server side, it significantly reduces communication overhead, making it a promising solution for improving data privacy in FL settings involving computationally limited IoT devices. To our knowledge, no complete HHE-FL framework has yet been proposed.

In this work, we propose a HHE framework for
FL that provides end-to-end privacy while remaining compatible
with resource-constrained settings. The main contributions of our work are the following:
the first end‑to‑end HHE‑FL framework that combines the PASTA stream cipher with the BFV FHE scheme, integrated into the Flower ecosystem; a single‑key distribution strategy and a homomorphic–evaluation routine that allows secure \emph{FedAvg} aggregation while keeping client encryption and communication costs minimal; a comprehensive experimental evaluation on Independent and Identically Distributed (IID) partitioned MNIST (12 clients, 10 rounds) that showed our HHE approach achieving 97.6\% accuracy (vs. 98.9\% in plaintext), while reducing client upload traffic by 2000$\times$. Server computational overhead increased by approximately 15,621$\times$ per client participating in the training phase, an issue to be addressed in future work.

The paper is organized as follows. Section~\ref{backgroudn} recalls foundational concepts. Section~\ref{sota} reviews the literature in FL with HE. Section~\ref{problemStatement} details our HHE-FL approach. Section~\ref{subsec:flower} discusses implementation. Section~\ref{results} presents and discusses results. Section~\ref{sc: conclusion} concludes with a summary, limitations, and future work.

\section{Backgound}\label{backgroudn}

This section presents the fundamentals of FL, including architectures, aggregation algorithms, and privacy risks. It overviews HE and most used schemes, and then introduces HHE, including a discussion on HE-friendly symmetric ciphers.

\subsection{Federated Learning and Privacy Threats}

\begin{figure}[hbt]
    \centering
\includegraphics[width=0.82\linewidth]{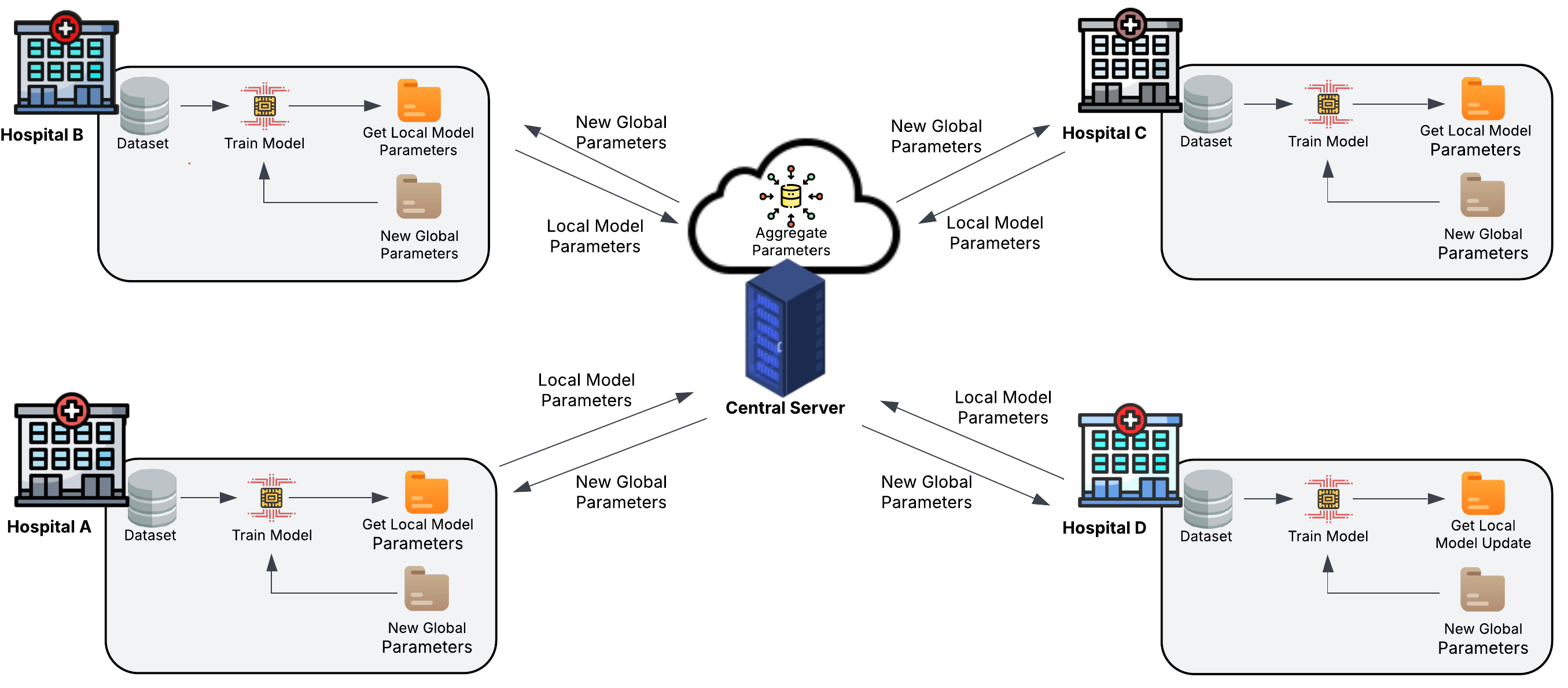}
    \caption{Federated Learning Architecture}

    \label{fig:fl-basic}
\end{figure}

FL is a distributed learning paradigm in which multiple clients collaboratively train a shared global model under the coordination of a central server, without directly exchanging their local data~\cite{mcmahan2023communicationefficientlearning}. Fig.~\ref{fig:fl-basic} illustrates a typical setup with four clients and one server. This setup represents one of the most common FL architectures, known as the \textit{centralized topology}, where a central server orchestrates the training process~\cite{liCentralizedDecentralizedFederated2025}.
Each client holds a local dataset and trains a model using their private data. After local training, the client sends an update, which is typically a model weight vector or gradient, to the server. The server then aggregates the updates from all selected clients to produce a new global model, which is sent back to the clients for the next training round.

\vspace{-0.3cm}    

\vspace{-0.4cm}

Federated Averaging~(FedAvg) is the most used aggregation algorithm, which performs weighted averaging of client updates based on local dataset sizes~\cite{mcmahan2023communicationefficientlearning}. Let \( K \) be the number of participating clients in a training round. For each client \( k \in \{1, \ldots, K\} \), let \( n_k \) be the number of local training samples, and let \( n = \sum_{k=1}^K n_k \) be the total number of data samples across all participating clients. Denote by \( w_{t+1}^k \) the updated model weights of client \( k \) after local training at round \( t+1 \). Then, the global model update is computed as:

\begin{equation}
w_{t+1} = \sum_{k=1}^K \frac{n_k}{n} w_{t+1}^k.
\end{equation}

A related variant, Federated Stochastic Gradient Descent (FedSGD), aggregates gradients computed on client data rather than updated model weights~\cite{mcmahan2023communicationefficientlearning}. 
In practice, FL deployments fall into two main categories:  
\emph{cross‑silo} FL, which involves a relatively small number of reliable institutional clients (e.g., hospitals or banks), and \emph{cross‑device} FL, which engages thousands to millions of resource‑constrained that are offline most of the time and connect only intermittently.  
Communication bandwidth, client drop‑outs, and limited computational resources are the main bottlenecks in the cross‑device setting~\cite{kairouzAdvancesOpenProblems2021,majeedCrossSiloModelBasedSecure2021}.   Since our objective is to make privacy‑preserving FL viable on such
devices, our proposed system targets this setting.

As mentioned before, FL enables collaboration with different clients while ensuring that data remains stored on individual devices. Even though raw data never leaves those devices, FL still has to deal with both external (e.g. eavesdroppers) and internal threats (e.g. malicious participants). A common internal adversary model is the
\emph{honest‑but‑curious} server, which follows the protocol correctly but
tries to infer private client information from the updates it receives.
Local model updates shared during FL retain subtle patterns that can leak private information. For instance, the server can reconstruct inputs by numerically inverting the shared gradients, perform a membership inference attack, or
test whether a specific record was in a client’s dataset~\cite{10089719,10.1007/978-981-97-5603-2_36,NIU2024404,Yu2023}.  
These vulnerabilities exist even when all parties follow the standard FedAvg
protocol, confirming that ``no raw data leaves the device'' is not a sufficient privacy guarantee.

\subsection{Homomorphic Encryption}

HE is a cryptographic primitive that allows computations over encrypted data, resulting in ciphertexts that, when decrypted, produce the same result as if the operations had been applied to the plaintexts~\cite{sathyaReviewHomomorphicEncryption2018}.
There are three main categories of HE schemes, namely Partial Homomorphic Encryption~(PHE), Some What Homomorphic Encryption (SWHE) and Fully Homomorphic Encryption~(FHE). These categories are based on how many and which operations are allowed over encrypted data: PHE schemes allows only one type of arithmetic operation (e.g. addition) an unlimited number of times; SWHE supports a limited set of operations, with restrictions on how many times they can be computed; FHE enables multiple types of operations to be performed an unlimited number of times. Due to the context of this work, we focus on the latter type.

Currently, there are four main FHE schemes, namely TFHE (Fast Fully Homomorphic Encryption over the Torus)~\cite{chillottiTFHEFastFully2018}, which operates at the bit level, BGV~\cite{brakerskiLeveledFullyHomomorphic} and BFV~\cite{fanSomewhatPracticalFully2012}, which are very similar and allow for exact computations over integers, and CKKS~\cite{cheonHomomorphicEncryptionArithmetic2016}, which supports approximate homomorphic computations over complex numbers.
However, since the security of HE schemes relies on complex mathematical problems, these schemes tend to be impractical, especially FHE schemes~\cite{gentryPaper}.
Two of the main issues that limit FHE from becoming a practical solution are \textit{ciphertext expansion} and \textit{noise growth}~\cite{marcolla}. The former refers to the growth, in size, that happens when encrypting a plaintext or after every computation performed, and the latter refers to the increase of noise after each operation, especially expensive ones such as multiplication.
This causes two major problems for implementation: ciphertext expansion makes communication, storage and computation much more expensive; noise growth limits the depth of computable circuits, since correct decryption can not be ensured if the noise grows beyond a certain threshold~\cite{dobraunigPastaCaseHybrid2021}.
New solutions, such as HHE, have been proposed to tackle these issues inherent to HE. The basic idea consists of combining HE and Symmetric Encryption~(SE), which is known for its fast encryption, transmission, and constant ciphertext expansion.

\subsection{HHE and HE-friendly ciphers}

HHE is a cryptographic concept that combines symmetric ciphers with HE to mitigate the significant computational and communication overhead traditionally associated with HE. Contrary to HE, SE schemes rely on a single key for both encryption and decryption, making them significantly faster.
In practice, HHE leverages the efficiency of SE for protecting data, while relying on FHE to enable secure computation over encrypted inputs. The idea dates back to the \textit{transciphering} framework introduced by Lauter et al.~\cite{lauterCanHomomorphicEncryption2011}, where a plaintext message \( m \) is encrypted using a symmetric cipher with key \( sk \), producing a lightweight ciphertext \( c_{{SKE}} \). The symmetric key \( sk \) is then encrypted under the homomorphic public key \( HE\_pk \), resulting in \( sk_{HE} \). Both \( c_{{SKE}} \) and \( sk_{HE} \) are sent to the server. The server then homomorphically evaluates the symmetric decryption circuit, a step referred in this work as \emph{Homomorphic Evaluation of Symmetric Decryption (HESD)}, to transform \( c_{{SKE}} \) into a homomorphic ciphertext \( m_{HE} \), without learning the plaintext.
 
In a client–server setting, HHE significantly reduces the computational burden on clients, since they only perform lightweight SE. It also lowers communication overhead, as symmetric ciphertexts are much smaller than their homomorphic equivalents. This makes HHE attractive for resource-constrained environments like the IoT, where devices often lack the capacity to support FHE directly.
The added server-side cost of homomorphic evaluation is a trade-off, but one that can be mitigated through the use of HE-friendly symmetric ciphers, which are specifically designed to minimize homomorphic computation by having low multiplicative depth.
As a result, they make \textit{transciphering}-based approaches more practical, especially in resource-constrained scenarios.

Most HE-friendly ciphers are designed for compatibility with one or more of the major FHE schemes: BGV, BFV, CKKS, and TFHE. Since BGV, BFV, and CKKS share similar algebraic structures, some ciphers can support multiple schemes. For example, compatibility with BGV and BFV is more natural for ciphers operating over integers, while CKKS supports approximate arithmetic over complex numbers.
Among the most prominent HE-friendly ciphers targeting BGV and BFV are RASTA~\cite{dobraunigRastaCipherLow2018}, DASTA~\cite{dasta}, and PASTA~\cite{dobraunigPastaCaseHybrid2021}. These designs are iterative improvements: DASTA refines RASTA, and PASTA further improves DASTA by reducing computational cost. In particular, PASTA is a stream cipher optimized for integer operations over \( \mathbb{F}_p \), with \( p \) a 16-bit prime. It is specifically designed to minimize homomorphic evaluation depth and has been shown to outperform earlier designs in terms of efficiency.

The authors of PASTA made available an open-source C++ framework\footnote{\url{https://github.com/isec-tugraz/hybrid-HE-framework}}, developed specially for benchmarking HHE schemes. It includes implementations of several HE-friendly ciphers, like PASTA, together with various HE schemes, such as BGV via HElib~\cite{haleviDesignImplementationHElib2020}, BFV via Microsoft SEAL~\cite{dowlinManualUsingHomomorphic2017}, and TFHE using the TFHE library\footnote{\url{https://github.com/tfhe/tfhe}}.

\vspace{-0.2cm}
\section{Related Work}\label{sota}

FL with HE has been extensively studied in the literature as a means of enabling privacy-preserving collaboration between distributed clients to train a global model. Existing work in this area explores various HE schemes to secure model updates. In this section, we review FL protocols that use HE, highlighting their architectural designs, aggregation strategies, and cryptographic mechanisms.

Several proposals choose PHE as their main HE component in the solution, ones employing FedAvg~\cite{rabieinejadTwoLevelPrivacyPreservingFramework2024,songSecureEfficientFederated2024,zhangPrivacyEAFLPrivacyEnhancedAggregation2023}, and others developing new aggregation algorithms, like Gradient Similarity Model Aggregation~\cite{wangNIDSFGPAFederatedLearning2024}.
Proposals relying on FHE often present centralized FL architectures, although some approaches, such as the work of Hijazi et al.~\cite{hijaziCollaborativeIoTLearning2024}, propose a decentralized solution where no server exists. Instead, a capable user is selected in each round to coordinate and aggregate local updates.
Their work combines techniques for data relevance, like cosine similarity, and clustering, to help prevent poisoning attacks. 
On the other hand, the centralized architectures can also be split between approaches that use FedAvg, or not, as their aggregation algorithm. 
Ma et al.~\cite{maPrivacypreservingFederatedLearning2022} proposed a centralized solution that uses FedAvg and introduced a variation of CKKS, namely xMK-CKKS, which allows the users to utilize distinct homomorphic keys. Here, the users compute a shared public key, which is used to encrypt the local model weights. To decrypt, users need to compute their decryption shares and send them to the server. Zhang et al.~\cite{zhangFaulttolerantFederatedLearning2024} proposes an improved version of the xMK-CKKS scheme, named ftxMK-CKKS, which provides fault tolerance when users drop out of the training process. They also encrypt local model weights and use FedAvg for aggregation.
The work of Duy et al.~\cite{duyFedChainHunterReliablePrivacypreserving2023} proposes a shared key distribution mechanism, but within a semi-decentralized FL framework by leveraging a blockchain and an aggregator server. To further protect client privacy, the authors incorporate DP.

Besides FedAvg, several other aggregation algorithms can be used. Tan et al.~\cite{tanPrivacypreservingFederatedLearning2024} proposed the use of a simple average algorithm in a centralized approach. To ensure the authenticity of the communicating parties, the authors use Transport Layer Security.
Fontela-Romero et al.~\cite{fontenla-romeroFedHEONNFederatedHomomorphically2023} proposed a centralized solution that introduces an Incremental Singular Value Decomposition Aggregation algorithm to reduce data dimensionality. The work of Xu et al.~\cite{xuEdgeServerEnhanced2024}, chose FedSGD as aggregation algorithm, using CKKS scheme to encrypt local gradients. It is also used Threshold Secret Sharing to make the scheme drop-out resilient. 

These works demonstrate that while HE, especially FHE, provides privacy guarantees, its use often depends on complementary techniques like secret sharing or DP, each introducing its own set of trade-offs. However, none of the surveyed studies applied HHE, leaving its suitability unexplored in FL.

\vspace{-0.2cm}
\section{Federated Learning with HHE}\label{problemStatement}

\begin{figure}[htb]
    \centering
    \includegraphics[width=\linewidth]{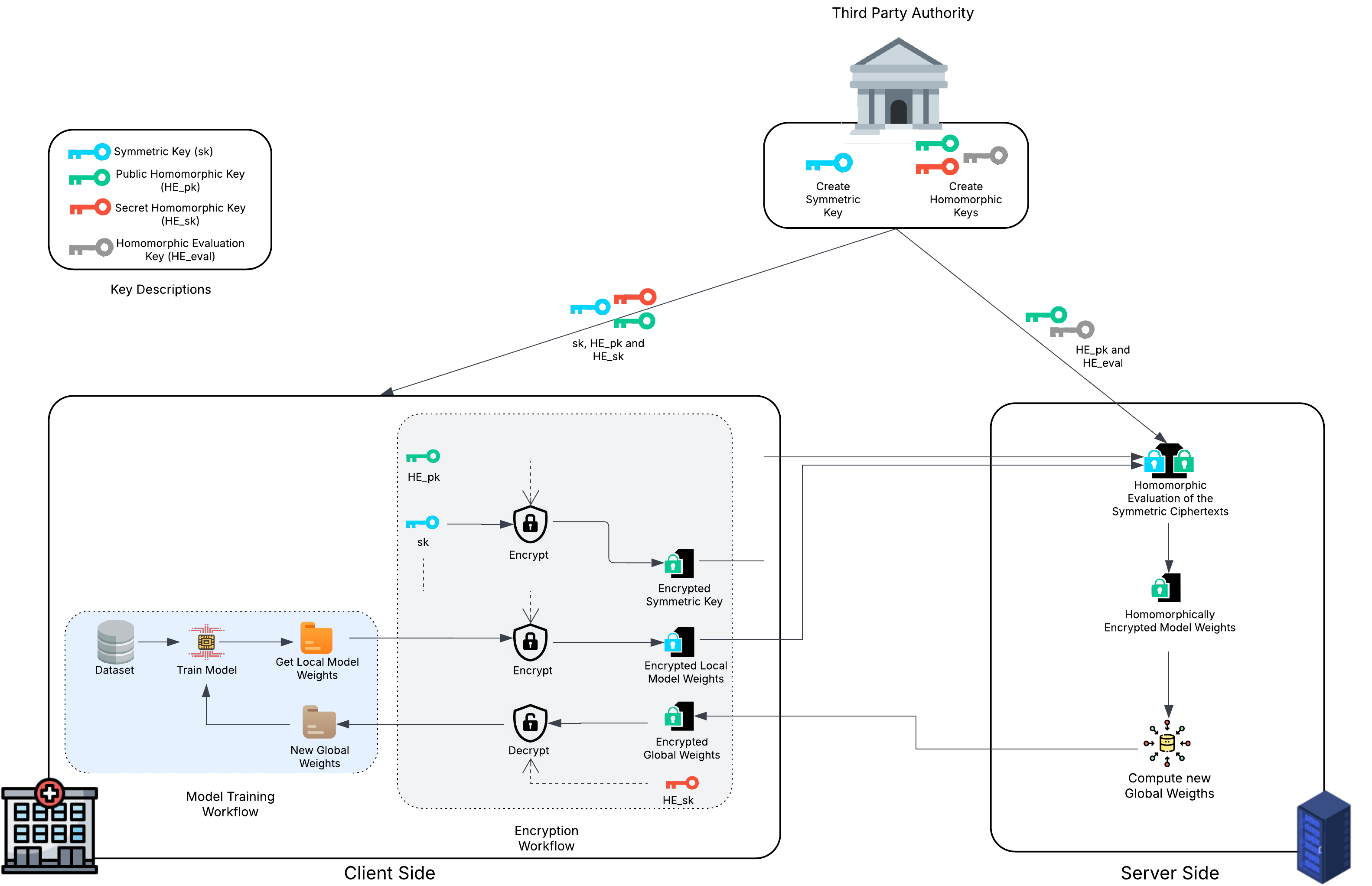}
    \caption{FL Architecture with HHE}
    \label{fig:hhe_with_eval}
\end{figure}

The main objective of this work is the creation of a privacy-preserving FL architecture for resource-constrained environments, such as those found in IoT settings. Our approach uses an HHE scheme that allows us to use a symmetric cipher to encrypt the data on the client side, while computationally intensive homomorphic operations are offloaded to the server.
Fig.~\ref{fig:hhe_with_eval} provides a high-level architecture of our proposed solution, considering a single client for illustration purposes. As shown, the proposed system involves three main entities: the \textit{Third-Party Authority} (TPA), the \textit{Server}, and the \textit{Clients}. The TPA is an independent trusted authority responsible for generating and distributing all cryptographic keys. The \textit{Server} is responsible for coordinating the FL training process, collecting encrypted updates from \textit{Clients}, performing secure aggregation, and distributing the updated global model. Finally, the \textit{Clients} train their local model, encrypt their local updates and symmetric key, send them to the \textit{Server}, and decrypt the received global model. In the following subsection, a detailed description of the operational workflow of our system is provided.

\subsection{Workflow of the proposed solution}\label{sect:workflow}
\begin{figure}[ht]
    \centering
    \includegraphics[width=0.66\linewidth]{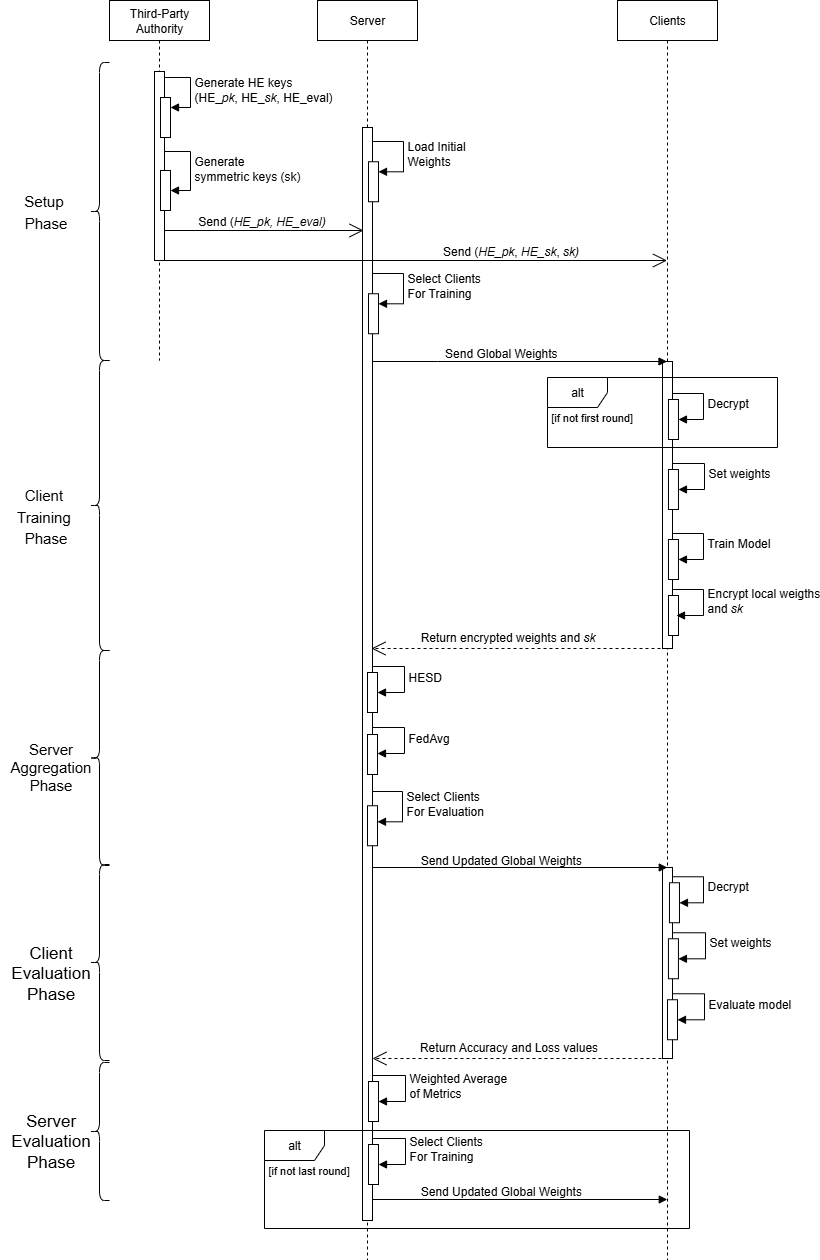}
    \caption{Workflow of the proposed solution}
    \label{fig:sequence_diagram}
    \vspace{-0.5cm}
\end{figure}

In Fig.~\ref{fig:sequence_diagram}, we illustrate the complete workflow of our proposed solution, which is comprised of five main phases: \textit{Setup}, \textit{Client Training}, \textit{Server Aggregation}, \textit{Client Evaluation}, and \textit{Server Evaluation}. In what follows, we describe all these phases.

\noindent
\textbf{Setup Phase}
In this phase, which occurs once at the beginning of our protocol, the TPA generates and distributes the necessary cryptographic keys to the clients and server. This includes a single tuple of homomorphic keys, (\textit{HE\_pk}, \textit{HE\_sk}, \textit{HE\_eval}), where \textit{HE\_pk} is the homomorphic public key, \textit{HE\_sk} is the homomorphic secret key, and \textit{HE\_eval} is the homomorphic evaluation key. Additionally, the TPA generates a unique symmetric key \(sk_i\) for each client \(i\), where \(1 \le i \le N\) and $N$ denotes the total number of clients in the system. The TPA then sends \((HE\_pk, HE\_eval)\) to the server, and \((HE\_pk, HE\_sk, sk_i)\) to each client \(i\). The server initializes a model and sends its weights to the clients, who use them as the starting point for local training during the training phase.

\smallskip
\noindent
\textbf{Client Training Phase} 
The \textit{Client Training Phase} begins after the \textit{Setup Phase} in the first round, with clients receiving the global model weights in plaintext. In subsequent rounds, it follows the \textit{Server Evaluation Phase},  where clients first decrypt the received encrypted global weights using their HE private key, \( HE\_sk \). Each client then trains its local model with the received weights, encrypts the update weights \(w_i\) with \(sk_i\), and encrypts \(sk_i\) with \(HE\_pk\). The resulting ciphertexts \({w^i_{SKE}}\) and \({sk^i_{HE}}\), together with the number of training samples \( n_i = \textit{training\_data}_i / batch\_size\), are sent to the server for aggregation.

\noindent
\textbf{Server Aggregation Phase}
The \textit{Server Aggregation Phase} occurs after every \textit{Client Training Phase} in each round of the FL process. During this phase, the server receives the \( w_{SKE} \) along with the corresponding homomorphically encrypted secret key \(sk_{HE}\) from the selected clients. The server then applies HESD on each \(w_{SKE}\), using the associated \(sk_{HE}\). Once all updates are transformed, the server aggregates the homomorphic ciphertexts to compute the new global model weights. In the current implementation, any client that drops out during the\textit{ Client Training Phase} is excluded from the aggregation for that round but may participate in subsequent rounds. Finally, the server selects a new subset of clients and sends them the updated model for evaluation.

\smallskip\noindent
\textbf{Client Evaluation Phase}
This phase follows the \textit{Server Aggregation Phase} in each round of the FL process. Selected clients receive the newly aggregated global weights and decrypt them using \textit{HE\_sk}. They update their local models with these weights and evaluate them on their local test dataset. Each client then sends the resulting accuracy and loss metrics to the server, along with the number of test samples.

\smallskip\noindent
\textbf{Server Evaluation Phase}
The \textit{Server Evaluation Phase} occurs after the \textit{Client Evaluation Phase} and concludes each round of the federated process. Upon receiving evaluation metrics from all clients, the server aggregates these metrics using a chosen aggregation algorithm. If additional training rounds remain, the server selects a new subset of clients for the upcoming \textit{Client Training Phase} and distributes the updated global weights accordingly.

All phases, except setup, are repeated for a fixed number of rounds.

\subsection{Threat Model}\label{sec:threat-model}

The security of the proposed solution is based on the following assumptions about the system setup and participants: \textit{A.1. The Setup Phase is trusted}: all cryptographic keys and the initial global weights are generated and distributed securely, with no adversarial interference; \textit{A.2. All clients are honest}: meaning they execute the protocol faithfully without attempting attacks; \textit{A.3. The server is honest-but-curious}:  it correctly executes the protocol but may attempt to infer private information; \textit{A.4. No collusion occurs between the server and clients; }\textit{A.5. External adversaries may attempt to infer information from participating clients but are unable to disrupt message transmission or inject poisoned data}.

Under these assumptions, the system provides the following security guarantees: \textit{SP.1. Confidentiality of individual client updates during transmission}; \textit{SP.2. Confidentiality of individual client updates during aggregation on the serve}; \textit{SP.3. Confidentiality of the aggregated model weights}.

\textit{SP.1} is ensured by client-side encryption. Each client encrypts its local model update with their symmetric key \textit{sk}, and \textit{sk} is then encrypted using the \(HE\_{pk}\). Therefore, the transmitted data remains confidential, preventing any eavesdropper from gaining access to the plaintext, even if they can observe the communication channel.
The server, upon receiving the encrypted client updates, performs HESD to generate the homomorphic ciphertexts used for the aggregation process. Since the updates remain encrypted throughout the process, and the server does not possess the \(HE\_{sk}\), it never has access to the plaintext of any individual client update, thereby ensuring \textit{SP.2.} Since aggregation is performed over homomorphic ciphertexts, the resulting aggregated model weights also remain encrypted. As the server does not possess the  \(HE\_sk\), it is unable to decrypt this aggregated result, thereby ensuring \textit{SP.3.} The framework provides no protection against malicious clients who may perform poisoning or inference attacks. A single malicious client could compromise confidentiality, as it could exploit access to the shared \( HE\_sk \) to decrypt the \( sk \) of an honest client, which is then used to decrypt that client's encrypted local update. Likewise, a malicious TPA, if introduced, could further endanger confidentiality by leaking \( HE\_sk \) to any malicious entity or disrupt the whole protocol by mismanaging key distribution.

\vspace{-0.3cm}
\section{Prototype Implementation} \label{subsec:flower} 
\vspace{-0.2cm}

\begin{figure}[h!]
  \centering
  \includegraphics[width=0.8\linewidth]{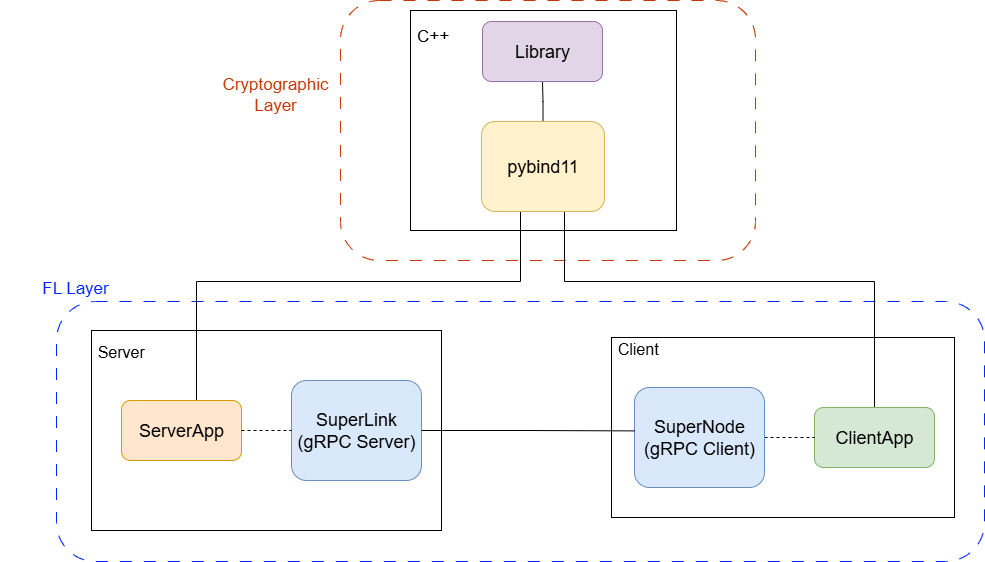}
  \caption{Prototype Implementation Architecture.}
  \label{fig:flower_arch}
  \vspace{-0.6cm}
\end{figure}

This section presents the prototype implementation of our solution. Fig.~\ref{fig:flower_arch} illustrates its overall architecture, in which two layers can be distinguished: the \textit{Cryptographic Layer} and the \textit{Flower Layer}.
Recall that our implementation focuses on \emph{cross‑device} FL scenarios in which resource‑constrained clients must encrypt their model updates with minimal overhead, while a well‑provisioned server performs the heavy homomorphic computations.  More specifically, we combine the PASTA stream cipher for SE, the BFV FHE scheme for server‑side processing, and the Flower FL framework for client-server orchestration, extended with a \texttt{C++} backend exposed to Python through \texttt{pybind11}\footnote{\url{https://github.com/pybind/pybind11}}. In what follows, we first detail each of the layers represented in Fig.~\ref{fig:flower_arch}, then describe the processing that occurs on the server side, followed by the processing on the client side.

\smallskip
\noindent
\textbf{Cryptographic Layer}
Our choice of an HE‑friendly SE scheme suitable for scenarios involving resource-constrained clients fell on PASTA~\cite{dobraunigPastaCaseHybrid2021} because its design enables low multiplicative depth and efficient homomorphic evaluation.   We paired it with BFV, which when compared to BGV, offers slightly simpler initial parameter selection as well as more straightforward implementation of common operations, such as encoding and noise management. It also allows integer plaintexts that align well with PASTA’s design.  
PASTA was integrated in our prototype using the authors' open-source \texttt{C++} framework\footnote{\url{https://github.com/isec-tugraz/hybrid-HE-framework}}, which also provides wrapper classes for BFV via Microsoft SEAL.

\noindent
\textbf{FL Layer}
Several FL frameworks were analyzed and compared based on learning modes, privacy-preserving mechanisms, encryption support, and scalability. Flower, proposed by Beutel et al.~\cite{beutelFlowerFriendlyFederated2022}, was selected for its high customizability, scalability, cloud compatibility, and the availability of a HE integration by Catalfamo et al.~\cite{catalfamoFlowerFullCompliantImplementation2024}. Since Flower is Python‑based and
the PASTA/BFV stack is implemented in \texttt{C++}, we expose the \texttt{C++} API via
\texttt{pybind11}.  
In Flower, the server consists of two main components: the \texttt{SuperLink}, which handles client communication via gRPC, and the \texttt{ServerApp}, which contains the orchestration logic for training, aggregation, and evaluation. On the client side, the \texttt{SuperNode} manages communication, while the \texttt{ClientApp} handles local training and evaluation.
To support cryptographic operations, both the \texttt{ServerApp} and \texttt{ClientApp} are connected to a shared \texttt{C++} layer. This layer consists of a \texttt{pybind11 bridge}, which exposes \texttt{C++} functions to Python, and the framework used for PASTA.

\smallskip\noindent
\textbf{Server‑Side Processing}
The server-side processing consists of two main steps: the HESD operations and federated averaging of the model updates. Algorithm~\ref{alg:hesd_chunks} shows the chunk‑based HESD procedure: each chunk of the symmetrically encrypted weights is converted into a BFV ciphertext before aggregation.  After HESD, FedAvg multiplies each weight vector by the client’s sample count \(n_k\) and sums the results homomorphically.  Division by the global total \(n\) is deferred to the clients, because BFV does not support exact division in ciphertext space.

\vspace{-0.5cm}

\begin{algorithm}[h!]
\caption{Chunk-Based HESD for a Single Client}
\label{alg:hesd_chunks}
\begin{algorithmic}[1]
\scriptsize
\Require Client's encrypted symmetric parameters $w_{SKE}$, Client's encrypted symmetric key $sk_{HE}$, BFV cipher instance \texttt{BFV\_cipher}

\State Switch to client's encrypted key
\State \quad \Call{\texttt{BFV\_cipher}.\texttt{set\_encrypted\_key}}{$sk_{HE}$} 

\State Initialize list for the homomorphically encrypted parameters:
\State \quad \texttt{he\_params} $\gets$ empty list

\For{each \texttt{chunk} in $w_{SKE}$}
    \State  $ \texttt{he\_chunk} \gets \texttt{BFV\_cipher.HESD(chunk)}$
    \State Append \texttt{he\_chunk} to \texttt{he\_params}
\EndFor

\State \Return \texttt{he\_params}
\end{algorithmic}
\end{algorithm}

\noindent
\textbf{Client‑Side Processing}
On each device, the processing begins by the quantization procedure and encrypting local model updates with PASTA. It ends by decrypting and applying the dequantization method to the aggregated global model. 
To encrypt model parameters with PASTA, float32 values in \([- \alpha, \alpha]\) are converted to int8 values in \([-2^{b-1}, 2^{b-1} - 1]\) using scale quantization, as described by Lang et al.~\cite{langComprehensiveStudyQuantization2024}, by clipping each \( x \in w_i \) and applying

{
\small
\[
quantized\_x = round\left( x * \frac{2^{b-1}}{\alpha}\right)
\]
}

\noindent with \( b = 8 \) for int8 precision. Since PASTA requires unsigned 64-bit inputs, the quantized int8 values are converted to uint64 before chunk-based encryption, as detailed in Algorithm~\ref{alg:encrypt}.
Upon receiving the aggregated ciphertexts, each client decrypts the received global weights in chunks using \(HE\_sk\). The decrypted uint64 values are converted to signed 64-bit integers (int64), restoring negative values by subtracting the modulus from any value greater than half the \textit{modulus}. Clients then apply the averaging step from FedAvg by dividing the aggregated parameters by \(n\) in float32 for accuracy. Finally, the results are dequantized by dividing by the quantization scale used during encryption (see Algorithm~\ref{alg:client_decrypt}).

\vspace{-0.5cm}
\begin{algorithm}[ht]
\caption{Quantize and Encrypt Local Model Weights}
\label{alg:encrypt}
\footnotesize
\begin{algorithmic}[1]
\scriptsize
\Require Model Weights $w_i$ (list of float32 arrays), Clipping Bound $\alpha$, Chunk Size \texttt{chunk\_size}, PASTA cipher instance \texttt{pasta\_cipher}

\State Flatten all parameters into a single vector:
\State \quad \texttt{flat\_params} $\gets$ concatenate(flatten($layer$) for $layer$ in $w_i$)
\State Clip each value $x$ in \texttt{flat\_params} to $[-\alpha, \alpha]$:
\State \quad \texttt{clipped} $\gets$ clip(\texttt{flat\_params}, $-\alpha, \alpha$)
\State Define scale factor for int8 quantization:
\State \quad \texttt{scale\_factor} $\gets \frac{2^{8-1}-1}{\alpha} = \frac{127}{\alpha}$
\State Quantize each clipped value:
\State \quad \texttt{quantized\_int8} $\gets$ round(\texttt{clipped} $\times$ \texttt{scale\_factor})
\State Convert quantized values to unsigned 64-bit integers:
\State \quad \texttt{quantized\_uint64} $\gets$ convert\_to\_uint64(\texttt{quantized\_int8})
\State Initialize list for encrypted chunks:
\State \quad \texttt{list\_sym\_ciphertext} $\gets$ empty list
\For{$i=0$ to length(\texttt{quantized\_uint64}) step \texttt{chunk\_size}}
    \State \texttt{chunk} $\gets$ \texttt{quantized\_uint64}$[i : i + \texttt{chunk\_size}]$
    \State \texttt{encrypted\_chunk} $\gets$ \texttt{pasta\_cipher.encrypt(chunk)}
    \State Append \texttt{encrypted\_chunk} to \texttt{list\_sym\_ciphertext}
\EndFor
\State \Return \texttt{list\_sym\_ciphertext}
\end{algorithmic}
\end{algorithm}

\vspace{-0.8cm}

\section{Experimental Evaluation and Analysis}\label{results}
\vspace{-0.2cm}

In this section, we present and analyze the results of our experimental evaluation of the proposed HHE-based FL architecture. We compare three setups using the MNIST dataset: a plaintext FL architecture, HHE-based FL combining PASTA and BFV, and FL using only BFV. In the two latter setups, clients share the same homomorphic keys. 
During training, accuracy, loss, communication, and computation costs are measured. Communication cost is defined as the total volume of data transmitted, while computation cost is the average CPU time required to perform an operation.
The final model is evaluated using additional metrics such as precision, recall, and F1-score.
\vspace{-0.2cm}

\begin{algorithm}[h!]
\caption{Decrypt and Dequantize Global Weights}
\label{alg:client_decrypt}
\footnotesize
\begin{algorithmic}[1]
\scriptsize
\Require Encrypted Global Weights $w_{HE}$, Decryption Key $HE\_sk$, Total Number of Examples $n$,  Clipping Bound $\alpha$, Chunk Size \texttt{chunk\_size}, Modulus \texttt{modulus} , BFV cipher instance \texttt{BFV\_cipher}

\State Initialize list for decrypted chunks:
\State \quad \texttt{decrypted\_chunks} $\gets$ empty 

\For{each \texttt{chunk} in $w_{HE}$}
    \State $c \gets \Call{\texttt{BFV\_cipher.decrypt}}{\texttt{chunk}, HE\_sk}$
    \State Append $c$ to \texttt{decrypted\_chunks}
\EndFor

\State Concatenate all chunks into one vector:
\State \quad \texttt{uint64\_params} $\gets$ concatenate(\texttt{decrypted\_chunks})

\State Convert to int64:
\State \quad \texttt{int64\_params} $\gets$ convert\_to\_int64(\texttt{uint64\_params})

\For{each $x$ in \texttt{int64\_params}}
    \If{$x > \texttt{modulus} \texttt{//}  2$}
        \State $x \gets x - \texttt{modulus}$
    \EndIf
\EndFor

\State Compute average:
\State \quad $\texttt{float\_params} \gets \texttt{int64\_params} \div n$

\State Define scale factor for dequantization:
\State \quad \texttt{scale\_factor} $\gets \frac{2^{8-1}-1}{\alpha} = \frac{127}{\alpha}$

\State Dequantize new global parameters:
\State \quad \texttt{decrypted\_params}  \( \gets \texttt{float\_params} \div \texttt{scale\_factor}\)

\State \Return \texttt{decrypted\_params}
\end{algorithmic}

\end{algorithm}

\subsection{ Experimental Setup}
\vspace{-0.2cm}

All the schemes were developed in Flower. The BFV-based FL architecture was implemented with the TenSEAL library, developed by Benaissa et al.~\cite{benaissaTenSEALLibraryEncrypted2021}, while the HHE prototype uses the C++ backend described in Section~\ref{subsec:flower}. All the experiments were run on Google Colab’s T4 High-RAM environment, which provided an NVIDIA T4 GPU, 51GB of system RAM, and 15GB of GPU memory.

The MNIST dataset, consisting of 70000 grayscale images of handwritten digits (28×28 pixels), was used, with 60000 samples for training and 10000 for testing. The training data was partitioned into $N$ client datasets in an IID manner using Flower's \texttt{IidPartitioner}. Each local dataset was then split into 80\% for training and 20\% for testing. Additionally, 20\% of each local training set was reserved for validation to enable early stopping during training rounds.

A Convolutional Neural Network~(CNN) based on the implementation by Gaurav Sharma~\cite{sharma_mnist_cnn_kaggle} was adopted for our experiments. It contains two convolutional layers, each followed by LeakyReLU activations and batch normalization. Max-pooling layers were used after each convolution for spatial downsampling. The resulting feature maps are flattened and sent through a dropout layer before the final dense layer generates predictions for 10 classes. The model contains approximately 8000 trainable parameters.

\subsection{Parameter Constraints and Configurations}\label{sect:limitations}
The HHE scheme used in this work operates over \(\mathbb{Z}_q\) with \(q = 2^{16} + 1\), requiring all server-side computations to remain within \([0, 2^{16} + 1[\). This bounds the product of the quantized model weights, the number of training clients, and the number of batches per client to be less than \(2^{16} + 1\). Given int8 quantization, this can be expressed as $ 2^8 \times 2^{x_1} \times 2^{x_2} < 2^{16} + 1$.
Thus, valid configurations must satisfy \(2^{x_1 + x_2} \leq 2^8\), where \(x_1\) is the number of batches per client and \(x_2\) the number of training clients. Only configurations estimated to complete within Colab’s 24-hour limit, based on per-round benchmarks, were executed.

\smallskip\noindent
\textbf{Chosen Configuration}\label{subsec:chosen}
Based on the constraints described before, our configuration consisted of 12 clients, each with 63 training batches, where 4 were used for training and 12 for evaluation, over 10 federated rounds. Table~\ref{tab:fl_parameters} lists all parameters used in the experiment. The value for $\alpha$ was chosen based on empirical observations of the model’s typical output range across several runs. 

\vspace{-0.8cm}
\begin{table}[ht]
\scriptsize
\centering
\caption{Parameters used for the FL experiment}
\label{tab:fl_parameters}
\begin{tabular}{l l |l p{1.1cm}}
\hline
Config. Name & Value & Config. Name & Value \\ \hline
Clients & 12 & Rounds & 10 \\
Epochs & 10 & Classifier & CNN \\
Loss function & Categorical \linebreak Cross-entropy & Batch size & 64 \\
Optimizer & Nadam & Learning Rate & 0.001 \\
Clients per Training Phase & 4 & Clients per Evaluation Phase & 12 \\
Clip Range ($\alpha$) & 5 & Key size & 256 bits \\
Plaintext size & 128 bits & Ciphertext size & 128 bits \\
Plaintext Modulus & 65537 & Polynomial Degree Modulus & 16384 \\
Security level & 128-bit & & \\
\hline
\end{tabular}
\end{table}

\vspace{-0.9cm}
\subsection{Experimental Results and Analysis}
\begin{figure}[ht]
    \centering
    \begin{minipage}[b]{0.48\linewidth}
        \includegraphics[width=\linewidth]{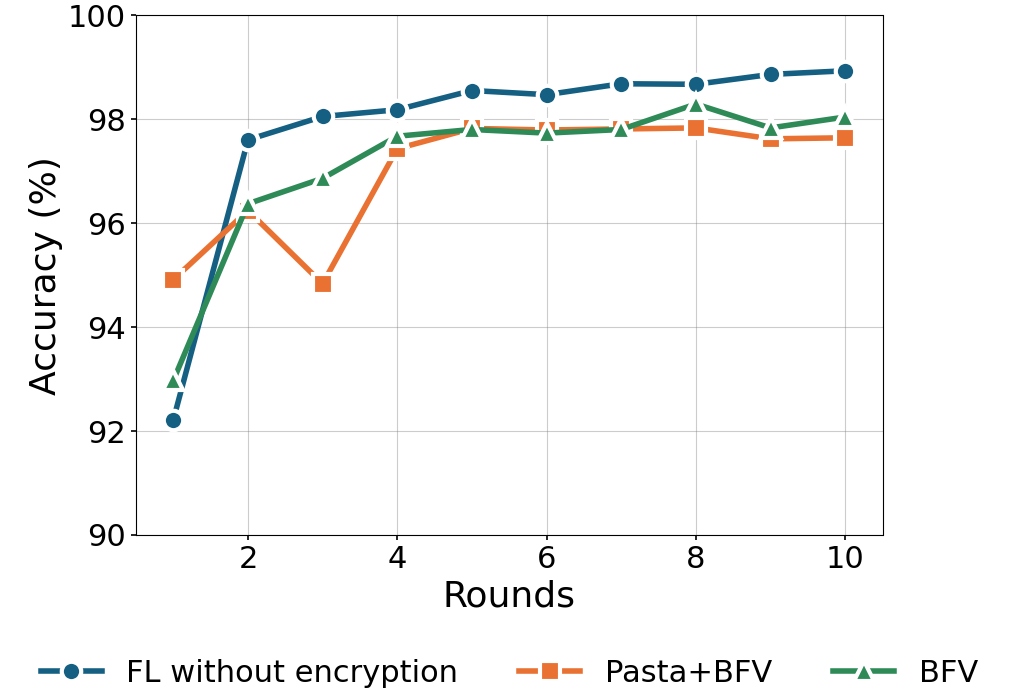}
    \end{minipage}
    \hfill
    \begin{minipage}[b]{0.48\linewidth}
        \includegraphics[width=\linewidth]{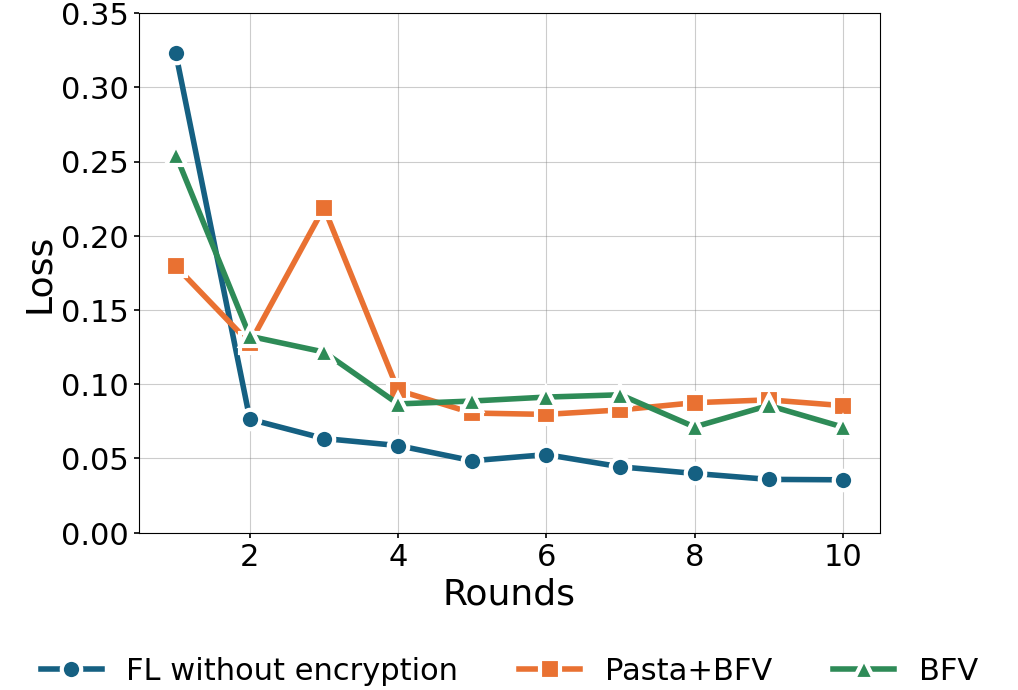}
    \end{minipage}
    \caption{Training Performance Across Rounds: (a) Accuracy; (b) Loss}
    \label{fig:accuracy_loss_fl}
    \vspace{-0.2cm}
\end{figure}

Regarding the models performance, Fig.~\ref{fig:accuracy_loss_fl} shows accuracy and loss over 10 rounds. It can be seen that the baseline plaintext FL converges to 98.93\% accuracy, BFV reaches 98.04\%, and
HHE attains 97.6\%, a less than 1.5\% drop attributable to quantization and HESD noise. Final precision, recall, and F1 in Table~\ref{tab:global_model_values} corroborate that the HHE-FL approach preserves model performance.

\vspace{-0.8cm}

\begin{table}[ht]
\scriptsize
\centering
\caption{Global Model Comparisons}
\resizebox{\linewidth}{!}{ 
\begin{tabular}{lccccc}
\hline
\multicolumn{1}{c}{} & Accuracy(\%) & Loss & Precision(\%) & Recall(\%) & F1-score(\%) \\ \hline
FL without encryption & 99.00 & 0.0360 & 99.00 & 99.00 & 99.00 \\ 
PASTA+BFV Scheme & 97.39 & 0.0962 & 97.50 & 97.39 & 97.39 \\ 
BFV Scheme & 98.21 & 0.0702 & 98.23 & 98.21 & 98.21 \\ \hline
\end{tabular}
}
\label{tab:global_model_values}
\end{table}

\begin{figure}[ht]
    \centering
    \begin{minipage}[b]{0.48\linewidth}
        \includegraphics[width=\linewidth]{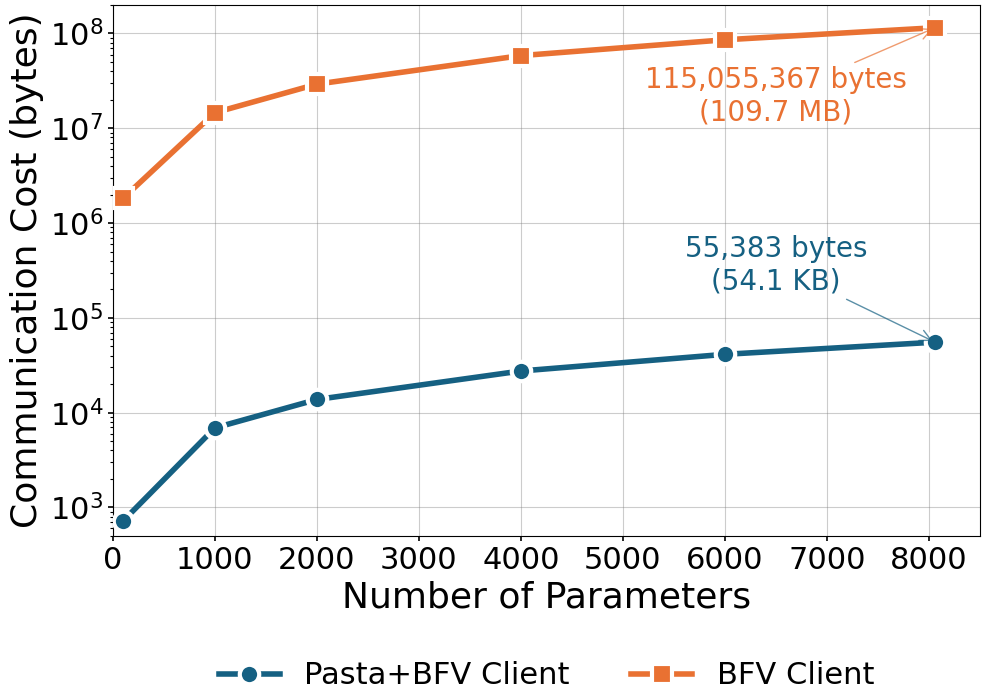}
    \end{minipage}
    \hfill
    \begin{minipage}[b]{0.48\linewidth}
        \includegraphics[width=\linewidth]{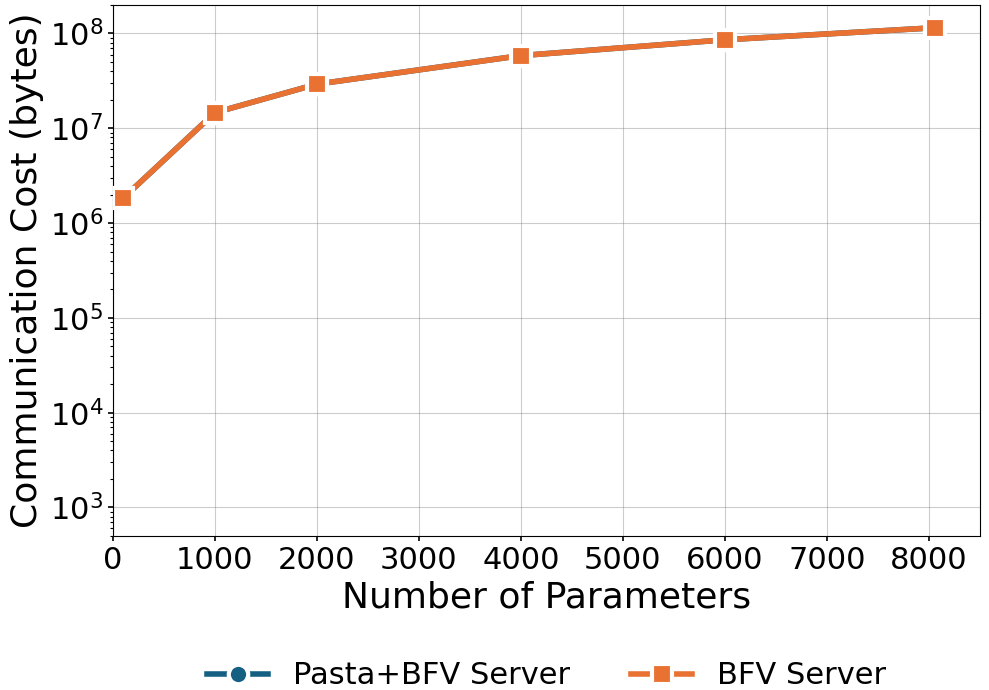}
        
    \end{minipage}
    \caption{ Communication cost comparison (values are shown on a logarithmic scale due to differences in magnitude): (a) From a single client to the server; (b) From the server to each client}
    \label{fig:client_computation_combined}
\vspace{-0.5cm}
\end{figure}

\vspace{-0.3cm}
The communication cost (Fig.~\ref{fig:client_computation_combined}a), measured has the amount of data sent between a single client and the server, shows that BFV results in over $2000\times$ higher upload cost compared to the proposed PASTA+BFV scheme, highlighting the benefit of using an hybrid approach and a cipher like PASTA. Fig.~\ref{fig:client_computation_combined}b shows that the amount of data sent from the server to each client is identical for both approaches. Notice that both rely on BFV on the server side. Overall, the PASTA+BFV scheme cuts total communication cost in half when compared to pure BFV, showing the improvement HHE brings in communication overhead.

Regarding computation cost, Fig.~\ref{fig:computation_cost_combined}a shows that using PASTA speeds up encryption and decryption by about $9.7\times$ and $6.9\times$, respectively. Training, taking  $\sim$10.7 seconds, dominates runtime and is identical across schemes since it runs in plaintext. Overall, PASTA+BFV’s total client runtime is $\sim$12.5 seconds, about 1.4× faster than BFV's $\sim$18 seconds, indicating that HHE offers a more efficient client-side solution than pure HE for resource-constrained IoT environments.

Although client-side computation differs by only 1.4×, server-side results show a larger gap. Fig.~\ref{fig:computation_cost_combined}b shows that server aggregation is significantly slower in the PASTA+BFV scheme, in part due to the high cost of homomorphic evaluation. Specifically, HESD for one client with 8000 parameters takes 1451 seconds, making the PASTA+BFV server with 4 clients about $62478\times$ slower than the BFV server for a single aggregation phase. These high runtimes are also impacted by the limitations of Google Colab, which runs on an Intel Xeon CPU @ 2.20GHz. In this environment, operations such as HESD were observed to be around $1.7\times$ slower than on our local machine (Intel Core i7-11370H, 16GB RAM).  In better environments, further time reductions are more likely to occur.

\vspace{-0.5cm}
\begin{figure}[htp]
    \centering
    \begin{minipage}[b]{0.48\linewidth}
        \centering
        \scriptsize
        \begin{tabular}{lcc}
        \hline
        {Operation} & {PASTA+BFV} & {BFV} \\ \hline
        Encryption (s)     & 0.32867  & 3.18009  \\
        Decryption (s)     & 0.42206  & 2.90076  \\
        Training (s)       & 10.70805 & 10.70805 \\
        Evaluation (s)     & 1.07516  & 1.07516  \\ \hline
        {Total runtime (s)} & 12.53394 & 17.86406 \\ \hline
        \end{tabular}
    \end{minipage}
    \hfill
    \begin{minipage}[b]{0.48\linewidth}
        \centering
        \scriptsize
        \begin{tabular}{lcc}
        \hline
        {Operation} & {PASTA+BFV} & {BFV} \\ \hline
        Aggregation (s)       & 0.2174     & 0.0929  \\
        HESD per client (s)   & 1451.0     & -       \\
        HESD per round (h)    & 1.6122     & -       \\ \hline
        {Total runtime (s)} & 5804.2174 & 0.0929 \\ \hline
        \end{tabular}
    \end{minipage}
    \caption{Average Computation Cost: (a) Client; (b) Server Aggregation Phase }
    \label{fig:computation_cost_combined}
    \vspace{-0.5cm}
\end{figure}

\begin{figure}[ht]
    \centering
    \includegraphics[width=0.48\linewidth]{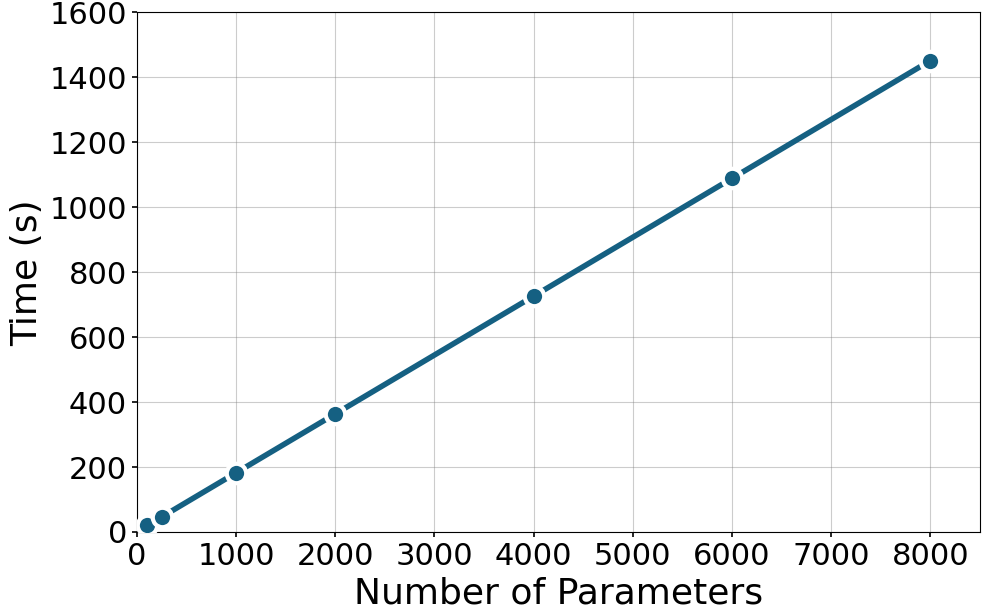}
    \caption{Runtime Overhead of HESD with Increasing Parameters}
    \label{fig:hesd_times}
\end{figure}

An analysis on how this cost scales with model size (Fig.~\ref{fig:hesd_times}) shows that the HESD time grows linearly with the number of parameters due to chunk-based processing and padding. If \(T_{\text{chunk}}\) is the time per chunk and \(N_{\text{chunks}}\) the total chunks, the total HESD time can be estimated as $
T_{\text{HESD}} = T_{\text{chunk}} \times N_{\text{chunks}}$.
During experiments, the BFV scheme experienced out-of-memory (OOM) failures starting at round 3, causing clients to drop out of training. These were due to Flower’s task manager, Ray, terminating processes when memory usage reached $\sim$48.5GB of 51GB (95\%). In contrast, the PASTA+BFV scheme completed all rounds without memory issues, demonstrating the memory-efficiency of HHE.

\vspace{-0.4cm}

\vspace{-1cm}

\section{Conclusion} \label{sc: conclusion}
\vspace{-0.1cm}

This work introduces the first HHE framework for FL that addresses privacy challenges in resource-constrained environments. By combining PASTA, a symmetric cipher, with BFV, a FHE scheme, the proposed architecture significantly reduces both client‑side computation overhead and communication cost.
Our experimental results showed that our hybrid approach maintains model performance when compared to a FHE approach while reducing client communication cost by 2000× and improving encryption and decryption speed by almost $10\times$. However, it increases server-side computation cost by $15621\times$ per client during the aggregation phase, due to the cost of the homomorphic evaluation of the PASTA decryption circuit and the limitations of Google Colab. Additionally, since BFV lacks real-number division, clients had to perform weighted averaging locally, posing a potential privacy issue.

Future work should aim to overcome current limitations. The addition of parallelization could help mitigate the server-side computational overhead. This parallelization could be applied at two levels: processing multiple clients concurrently, or processing the chunks of ciphertext within each client in parallel. Designing HE-friendly ciphers with multi-key support would improve security against collusion attacks, and implementing dropout-resilient mechanisms, such as secret sharing, would improve robustness during training and even help against poisoning attacks.  

\begin{credits}
\subsubsection{\ackname} This work has been supported by the PC2phish project, which has received funding from FCT with Refª: 2024.07648.IACDC. Furthermore, this work also received funding from the project UIDB/00760/2020.
\end{credits}

\bibliographystyle{splncs04}
\bibliography{BIB}

\begin{thebibliography}{10}
\providecommand{\url}[1]{\texttt{#1}}
\providecommand{\urlprefix}{URL }
\providecommand{\doi}[1]{https://doi.org/#1}

\bibitem{abdinasibfarHHELandExploring2025}
Abdinasibfar, H., Nuoskala, C., Michalas, A.: The {HHE} land: Exploring the landscape of hybrid homomorphic encryption. Cryptology {ePrint} Archive, Paper 2025/071 (2025), \url{https://eprint.iacr.org/2025/071}

\bibitem{aledhariFederatedLearningSurvey2020}
Aledhari, M., Razzak, R., Parizi, R.M., Saeed, F.: Federated {{Learning}}: {{A Survey}} on {{Enabling Technologies}}, {{Protocols}}, and {{Applications}}. IEEE Access  \textbf{8},  140699--140725 (2020). \doi{10.1109/ACCESS.2020.3013541}

\bibitem{benaissaTenSEALLibraryEncrypted2021}
Benaissa, A., Retiat, B., Cebere, B., Belfedhal, A.E.: {{TenSEAL}}: {{A Library}} for {{Encrypted Tensor Operations Using Homomorphic Encryption}} (Apr 2021). \doi{10.48550/arXiv.2104.03152}

\bibitem{beutelFlowerFriendlyFederated2022}
Beutel, D.J., Topal, T., Mathur, A., Qiu, X., {Fernandez-Marques}, J., Gao, Y., Sani, L., Li, K.H., Parcollet, T., de~Gusm{\~a}o, P.P.B., Lane, N.D.: Flower: {{A Friendly Federated Learning Research Framework}} (Mar 2022). \doi{10.48550/arXiv.2007.14390}

\bibitem{brakerskiLeveledFullyHomomorphic}
Brakerski, Z., Gentry, C., Vaikuntanathan, V.: Fully homomorphic encryption without bootstrapping. Cryptology {ePrint} Archive, Paper 2011/277 (2011), \url{https://eprint.iacr.org/2011/277}

\bibitem{catalfamoFlowerFullCompliantImplementation2024}
Catalfamo, A., Carnevale, L., Garofalo, M., Villari, M.: Flower {{Full-Compliant Implementation}} of {{Federated Learning}} with {{Homomorphic Encryption}}. In: 2024 {{IEEE Symposium}} on {{Computers}} and {{Communications}} ({{ISCC}}). pp.~1--5 (Jun 2024). \doi{10.1109/ISCC61673.2024.10733641}

\bibitem{cheonHomomorphicEncryptionArithmetic2016}
Cheon, J.H., Kim, A., Kim, M., Song, Y.: Homomorphic encryption for arithmetic of approximate numbers. In: Takagi, T., Peyrin, T. (eds.) Advances in Cryptology -- ASIACRYPT 2017. pp. 409--437. Springer International Publishing, Cham (2017). \doi{10.1007/978-3-319-70694-8_15}

\bibitem{chillottiTFHEFastFully2018}
Chillotti, I., Gama, N., Georgieva, M., Izabach{\`e}ne, M.: {TFHE}: Fast fully homomorphic encryption over the torus. J. Cryptology  \textbf{33}(1),  34--91 (Jan 2020). \doi{10.1007/s00145-019-09319-x}

\bibitem{dobraunigRastaCipherLow2018}
Dobraunig, C., Eichlseder, M., Grassi, L., Lallemand, V., Leander, G., List, E., Mendel, F., Rechberger, C.: Rasta: A cipher with low anddepth and few ands per bit. In: Shacham, H., Boldyreva, A. (eds.) Advances in Cryptology -- CRYPTO 2018. pp. 662--692. Springer International Publishing, Cham (2018). \doi{10.1007/978-3-319-96884-1_22}

\bibitem{dobraunigPastaCaseHybrid2021}
Dobraunig, C., Grassi, L., Helminger, L., Rechberger, C., Schofnegger, M., Walch, R.: Pasta: A case for hybrid homomorphic encryption. Cryptology {ePrint} Archive, Paper 2021/731 (2021), \url{https://eprint.iacr.org/2021/731}

\bibitem{dowlinManualUsingHomomorphic2017}
Dowlin, N., {Gilad-Bachrach}, R., Laine, K., Lauter, K., Naehrig, M., Wernsing, J.: Manual for {{Using Homomorphic Encryption}} for {{Bioinformatics}}. Proceedings of the IEEE pp. 1--16; (2017). \doi{10.1109/JPROC.2016.2622218}

\bibitem{duyFedChainHunterReliablePrivacypreserving2023}
Duy, P.T., Quyen, N.H., Khoa, N.H., Tran, T.D., Pham, V.H.: {{FedChain-Hunter}}: {{A}} reliable and privacy-preserving aggregation for federated threat hunting framework in {{SDN-based IIoT}}. Internet of Things  \textbf{24},  100966 (Dec 2023). \doi{10.1016/j.iot.2023.100966}

\bibitem{fanSomewhatPracticalFully2012}
Fan, J., Vercauteren, F.: Somewhat practical fully homomorphic encryption. Cryptology {ePrint} Archive, Paper 2012/144 (2012), \url{https://eprint.iacr.org/2012/144}

\bibitem{fontenla-romeroFedHEONNFederatedHomomorphically2023}
{Fontenla-Romero}, O.e.a.: {{FedHEONN}}: {{Federated}} and homomorphically encrypted learning method for one-layer neural networks. Future Generation Computer Systems  \textbf{149},  200--211 (Dec 2023). \doi{10.1016/j.future.2023.07.018}

\bibitem{gentryPaper}
Gentry, C.: Fully homomorphic encryption using ideal lattices. In: Proceedings of the 41st Annual ACM Symposium on Theory of Computing. p. 169–178. STOC '09, Association for Computing Machinery, New York, NY, USA (2009). \doi{10.1145/1536414.1536440}

\bibitem{haleviDesignImplementationHElib2020}
Halevi, S., Shoup, V.: Design and implementation of {HElib}: a homomorphic encryption library. Cryptology {ePrint} Archive, Paper 2020/1481 (2020), \url{https://eprint.iacr.org/2020/1481}

\bibitem{dasta}
Hebborn, P., Leander, G.: Dasta – alternative linear layer for rasta. IACR Transactions on Symmetric Cryptology  \textbf{2020}(3),  46–86 (Sep 2020). \doi{10.13154/tosc.v2020.i3.46-86}

\bibitem{hijaziCollaborativeIoTLearning2024}
Hijazi, N.M., Aloqaily, M., Guizani, M.: Collaborative {{IoT}} learning with secure peer-to-peer federated approach. Computer Communications  \textbf{228},  107948 (Dec 2024). \doi{10.1016/j.comcom.2024.107948}

\bibitem{jin2023fedmlhe}
Jin, W., Yao, Y., Han, S., Joe-Wong, C., Ravi, S., Avestimehr, S., He, C.: Fed{ML}-{HE}: An efficient homomorphic-encryption-based privacy-preserving federated learning system. In: International Workshop on Federated Learning in the Age of Foundation Models in Conjunction with NeurIPS 2023 (2023), \url{https://openreview.net/forum?id=PuYD0fh5aq}

\bibitem{kairouzAdvancesOpenProblems2021}
Kairouz, P., McMahan, H.B., et~al.: Advances and {{Open Problems}} in {{Federated Learning}} (Mar 2021). \doi{10.48550/arXiv.1912.04977}

\bibitem{langComprehensiveStudyQuantization2024}
Lang, J., Guo, Z., Huang, S.: A comprehensive study on quantization techniques for large language models (Oct 2024). \doi{10.48550/arXiv.2411.02530}

\bibitem{liCentralizedDecentralizedFederated2025}
Li, Q., Yu, W., Xia, Y., Pang, J.: From {{Centralized}} to {{Decentralized Federated Learning}}: {{Theoretical Insights}}, {{Privacy Preservation}}, and {{Robustness Challenges}} (Mar 2025). \doi{10.48550/arXiv.2503.07505}

\bibitem{maPrivacypreservingFederatedLearning2022}
Ma, J., Naas, S.A., Sigg, S., Lyu, X.: Privacy-preserving federated learning based on multi-key homomorphic encryption. International Journal of Intelligent Systems  \textbf{37}(9),  5880--5901 (2022). \doi{10.1002/int.22818}

\bibitem{majeedCrossSiloModelBasedSecure2021}
Majeed, U., Hassan, S.S., Hong, C.S.: Cross-{{Silo Model-Based Secure Federated Transfer Learning}} for {{Flow-Based Traffic Classification}}. In: 2021 {{International Conference}} on {{Information Networking}} ({{ICOIN}}). pp. 588--593. IEEE, Jeju Island, Korea (South) (Jan 2021). \doi{10.1109/ICOIN50884.2021.9333905}

\bibitem{marcolla}
Marcolla, C., Sucasas, V., Manzano, M., Bassoli, R., Fitzek, F.H.P., Aaraj, N.: Survey on fully homomorphic encryption, theory, and applications. Proceedings of the IEEE  \textbf{110}(10),  1572--1609 (2022). \doi{10.1109/JPROC.2022.3205665}

\bibitem{mcmahan2023communicationefficientlearning}
McMahan, H.B., E, M., D, R., S., H., y~Arcas, B.A.: Communication-efficient learning of deep networks from decentralized data (2023), \url{https://arxiv.org/abs/1602.05629}

\bibitem{lauterCanHomomorphicEncryption2011}
Naehrig, M., Lauter, K., Vaikuntanathan, V.: Can homomorphic encryption be practical? In: Proceedings of the 3rd ACM Workshop on Cloud Computing Security Workshop. p. 113–124. CCSW '11, Association for Computing Machinery, New York, NY, USA (2011), \url{https://doi.org/10.1145/2046660.2046682}

\bibitem{NIU2024404}
Niu, J., Liu, P., et~al.: A survey on membership inference attacks and defenses in machine learning. Journal of Information and Intelligence  \textbf{2}(5),  404--454 (2024). \doi{10.1016/j.jiixd.2024.02.001}

\bibitem{10089719}
Ovi, P.R., Gangopadhyay, A.: A comprehensive study of gradient inversion attacks in federated learning and baseline defense strategies. In: 2023 57th Annual Conference on Information Sciences and Systems (CISS). pp.~1--6 (2023). \doi{10.1109/CISS56502.2023.10089719}

\bibitem{10.1007/978-981-97-5603-2_36}
Qian, J., Wei, K., Wu, Y., Zhang, J., Chen, J., Bao, H.: Gi-smn: Gradient inversion attack against federated learning without prior knowledge. In: Huang, D.S., Chen, W., Pan, Y. (eds.) Advanced Intelligent Computing Technology and Applications. pp. 439--448. Springer Nature Singapore, Singapore (2024). \doi{10.1007/978-981-97-5603-2_36}

\bibitem{rabieinejadTwoLevelPrivacyPreservingFramework2024}
Rabieinejad, E., Yazdinejad, A., Dehghantanha, A., Srivastava, G.: Two-{{Level Privacy-Preserving Framework}}: {{Federated Learning}} for {{Attack Detection}} in the {{Consumer Internet}} of {{Things}}. IEEE Transactions on Consumer Electronics  \textbf{70}(1),  4258--4265 (Feb 2024). \doi{10.1109/TCE.2024.3349490}

\bibitem{sathyaReviewHomomorphicEncryption2018}
Sathya, S.S., Vepakomma, P., Raskar, R., Ramachandra, R., Bhattacharya, S.: A {{Review}} of {{Homomorphic Encryption Libraries}} for {{Secure Computation}} (Dec 2018). \doi{10.48550/arXiv.1812.02428}

\bibitem{sharma_mnist_cnn_kaggle}
Sharma, G.: Mnist cnn with 8k parameters (2020), \url{https://www.kaggle.com/code/gauravsharma99/mnist-8k-parameters}, accessed: 2025-06-11

\bibitem{songSecureEfficientFederated2024}
Song, C., Wang, Z., Peng, W., Yang, N.: Secure and {{Efficient Federated Learning Schemes}} for {{Healthcare Systems}}. Electronics  \textbf{13}(13), ~2620 (Jan 2024). \doi{10.3390/electronics13132620}

\bibitem{stanSecureFederatedLearning2022}
Stan, O., Thouvenot, V., Boudguiga, A., Kapusta, K., Zuber, M., Sirdey, R.: A {{Secure Federated Learning}}: Analysis of different cryptographic tools. In: {{SECRYPT}} 2022 - 19th {{International Conference}} on {{Security}} and {{Cryptography}}. vol.~1, p.~669 (Jul 2022). \doi{10.5220/0011322700003283}

\bibitem{tanPrivacypreservingFederatedLearning2024}
Tan, Z.S., {See-To}, E.W., Lee, K.Y., Dai, H.N., Wong, M.L.: Privacy-preserving federated learning for proactive maintenance of {{IoT-empowered}} multi-location smart city facilities. Journal of Network and Computer Applications  \textbf{231},  103996 (Nov 2024). \doi{10.1016/j.jnca.2024.103996}

\bibitem{wangNIDSFGPAFederatedLearning2024}
Wang, J., Yang, K., Li, M.: Nids-fgpa: A federated learning network intrusion detection algorithm based on secure aggregation of gradient similarity models. PLOS ONE  \textbf{19} (10 2024). \doi{10.1371/journal.pone.0308639}

\bibitem{xuEdgeServerEnhanced2024}
Xu, Y., Mao, Y., Li, J., Chen, X., Wu, S.: Edge server enhanced secure and privacy preserving federated learning. Computer Networks  \textbf{249},  110465 (Jul 2024). \doi{10.1016/j.comnet.2024.110465}

\bibitem{Yu2023}
Yu, S., Cui, L.: Inference Attacks and Counterattacks in Federated Learning, pp. 13--36. Springer Nature Singapore, Singapore (2023). \doi{10.1007/978-981-19-8692-5}

\bibitem{zhangHomomorphicEncryptionBasedPrivacyPreserving2023}
Zhang, L., Xu, J., Vijayakumar, P., Sharma, P.K., Ghosh, U.: Homomorphic {{Encryption-Based Privacy-Preserving Federated Learning}} in {{IoT-Enabled Healthcare System}}. IEEE Transactions on Network Science and Engineering  \textbf{10}(5),  2864--2880 (Sep 2023). \doi{10.1109/TNSE.2022.3185327}

\bibitem{zhangPrivacyEAFLPrivacyEnhancedAggregation2023}
Zhang, M., Chen, S., Shen, J., Susilo, W.: {{PrivacyEAFL}}: {{Privacy-Enhanced Aggregation}} for {{Federated Learning}} in {{Mobile Crowdsensing}}. IEEE Transactions on Information Forensics and Security  \textbf{18},  5804--5816 (2023). \doi{10.1109/TIFS.2023.3315526}

\bibitem{zhangFaulttolerantFederatedLearning2024}
Zhang, Y., Zhang, W., Shen, C.: A fault-tolerant federated learning scheme based on multi-key homomorphic encryption. In: Proceedings of the 2024 {{International Academic Conference}} on {{Edge Computing}}, {{Parallel}} and {{Distributed Computing}}. pp. 96--101. {{ECPDC}} '24, Association for Computing Machinery, New York, NY, USA (Aug 2024). \doi{10.1145/3677404.3677421}

\end{thebibliography}

\end{document}